\documentclass[letterpaper, 10 pt, conference]{ieeeconf}  % Comment this line out
\IEEEoverridecommandlockouts                              % This command is only
\usepackage{graphicx}
\usepackage{amsmath} % assumes amsmath package installed
\usepackage{rotating}
\usepackage{tabularray}
\usepackage{xcolor}
\usepackage{hyperref}
\usepackage{cleveref}
\usepackage{caption}
\usepackage{makecell}

\title{\LARGE \bf
    FairSCOSCA: Fairness At Arterial Signals --- Just Around The Corner
}

\author{Kevin Riehl, Justin Weiss, Anastasios Kouvelas and Michail A. Makridis% <-this % stops a space
\thanks{K. Riehl, J. Weiss, A. Kouvelas, and M. Makridis are with Traffic Engineering Group, Institute for Transportation Planning and Systems,
        ETH Zürich, Stefano Franscini Platz 3, 8053 Zürich, Switzerland
        {\tt\small kriehl@ethz.ch, juweiss@ethz.ch, kouvelas@ethz.ch, mmakridis@ethz.ch }.}%
}

\begin{document}

\maketitle
\thispagestyle{empty}
\pagestyle{empty}

%%%%%%%%%%%%%%%%%%%%%%%%%%%%%%%%%%%%%%%%%%%%%%%%%%%%%%%%%%%%%%%%%%%%%%%%%%%%%%%%
\begin{abstract}
Traffic signal control at intersections, especially in arterial networks, is a key lever for mitigating the growing issue of traffic congestion in cities. 
Despite the widespread deployment of SCOOTS and SCATS, which prioritize efficiency, fairness has remained largely absent from their design logic, often resulting in unfair outcomes for certain road users, such as excessive waiting times.
Fairness however, is a major driver of public acceptance for implementation of new controll systems.
Therefore, this work proposes FairSCOSCA, a fairness-enhancing extension to these systems, featuring two novel yet practical design adaptations grounded in multiple normative fairness definitions: (1) green phase optimization incorporating cumulative waiting times, and (2) early termination of underutilized green phases. Those extensions ensure fairer distributions of green times.
Evaluated in a calibrated microsimulation case study of the arterial network in Esslingen am Neckar (Germany), FairSCOSCA demonstrates substantial improvements across multiple fairness dimensions (Egalitarian, Rawlsian, Utilitarian, and Harsanyian) without sacrificing traffic efficiency. 
Compared against Fixed-Cycle, Max-Pressure, and standard SCOOTS/SCATS controllers, FairSCOSCA significantly reduces excessive waiting times, delay inequality and horizontal discrimination between arterial and feeder roads. 
This work contributes to the growing literature on equitable traffic control by bridging the gap between fairness theory and the practical enhancement of globally deployed signal systems.
Open source implementation available on GitHub.
\end{abstract}

%%%%%%%%%%%%%%%%%%%%%%%%%%%%%%%%%%%%%%%%%%%%%%%%%%%%%%%%%%%%%%%%%%%%%%%%%%%%%%%%
\section{INTRODUCTION}
Roads were built to serve the people by enabling mobility.
The rising traffic demand of the recent decades increasingly congests highways and cities, limiting the utility of driving for road users.
Traffic congestion is a problem causing externalities such as wasted time, emissions, fuel, noise, and psychological stress, leading to consequences for environment, public health, economy, and life quality~\cite{knieps2014congestion}.
According to the 2024 Global Traffic Scorecard published by INRIX, Inc. (leading authority in transportation analytics) average drivers in the most congested cities globally (Istanbul, New York, Chicago, London, Mexico City) lost more than 100 hours per year. In the United States alone, there is an estimated, economic damage of \$771 per driver annually~\cite{inrix2024scorecard}.
% •	Roads were build to serve the people, increasing traffic demand increasingly causes congestion and limits the utility of roads for users.
% •	Traffic congestion is a problem, externalities wasted time, emissions, noise, perceived stress, consequences for environment, public health, economic damage, life quality. 
The major bottleneck of roads in the city are intersections, which can be addressed with signalized intersection management through traffic light control.
Especially arterial networks, where busy arterial roads intersect with feeder networks are often subject to municipal investments. And it is those networks in particular, where it pays off to invest into sensors and actuated, controlled, traffic-responsive traffic lights rather than to rely on sensor-less, pre-timed signals~\cite{glukharev2013traffic,stoilova2024urban}.
% •	Congestion in cities major bottleneck is intersections, that can be addressed with signalized intersection management through traffic light control. Especially arterial networks, where busy arterial roads intersect with feeder networks, are often subject to municipal investments and control by placing sensors and traffic lights, there it pays off to invest in sensors to enable not only pretimed but actual real-time responsive traffic lights.

Various algorithms and techniques have been researched and explored for traffic light control, ranging from Fixed-Time~\cite{van2006delay}, Max-Pressure~\cite{varaiya2013max} to optimization~\cite{teo2010optimization} and machine learning based approaches~\cite{shirasaka2023distributed}.
In practice, the Sydney Coordinated Adaptive Traffic System (SCATS)~\cite{lowrie1990scats} and Split, Cycle, and Offset Optimization Technique for Signals (SCOOTS)~\cite{hunt1981scoot} count amongst the most established traffic light control systems. 
SCOOTS is leading the market in more than 350 cities globally since the 1980s~\cite{TRL_SCOOT}, and SCATS is used in 216 cities and 32 countries worldwide since 1975~\cite{SCATS_Home}.
Together, these systems control more than 60,000 intersections and 565 cities worldwide, influencing traffic flow and life reality for millions of vehicles day by day~\cite{SCATS_Home}.
% •	Various algorithms and techniques have been researched and explored for traffic light control, ranging from fixed time, max pressure, to … reinforcement learning. In practice, the most established algorithms are SCOOTS and SCATS (give numbers how many in the world)

\begin{figure*}[!ht]
    \centering
    \includegraphics[width=\linewidth]{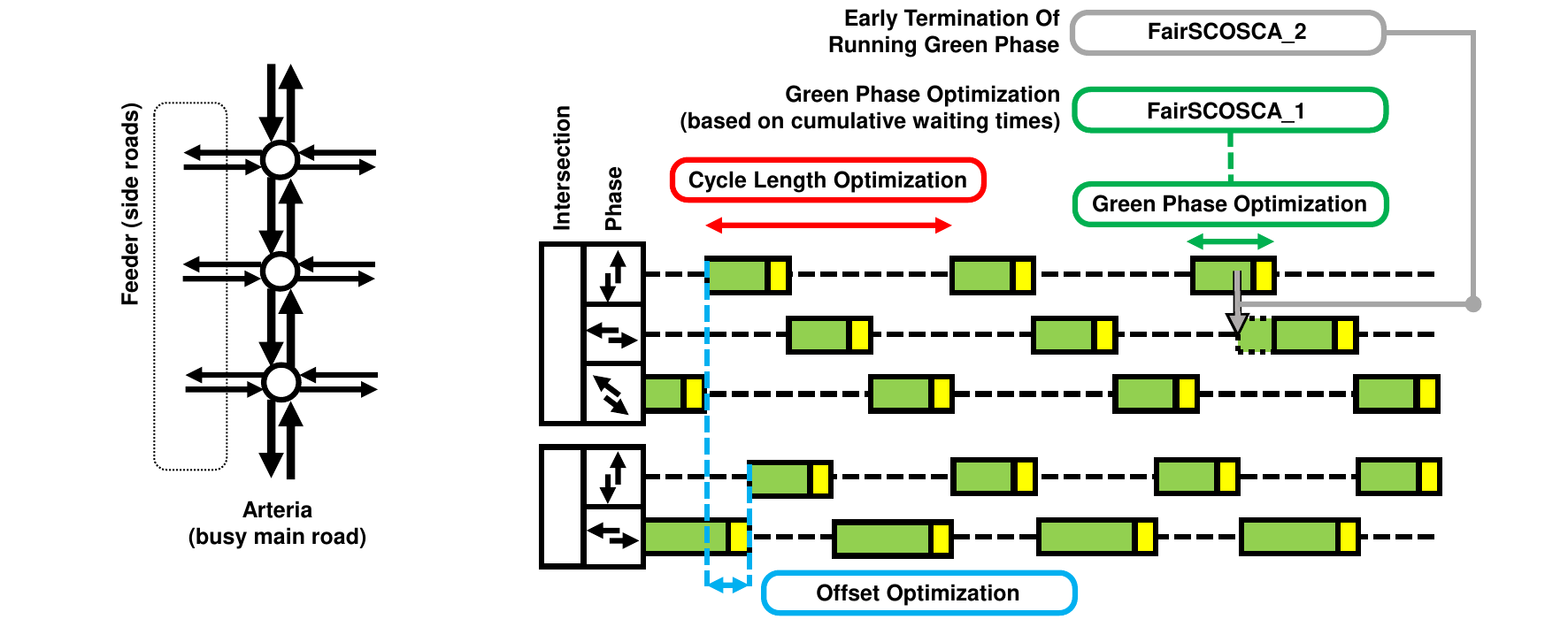}
    \caption{\textbf{Signalized Intersection Management with SCOOTS/SCATS:} SCOOTS/SCATS consist of three optimizers, that dynamically adjust cycle length, green phases, and offsets of traffic lights across an arterial road. The two proposed design features FairSCOSCA\_1 and FairSCOSCA\_2 are dedicated to improving the green phase optimization and to enable early termination of unused, running green phases. Doing so, the proposed methods can achieve higher levels of fairness without affecting efficiency too much. This visual illustrates the minimal yet impactful modifications needed to enhance fairness in a real-world-deployable manner.}
    \label{fig:scosca}
\end{figure*}

Prior works mainly focused on optimizing for traffic efficiency measures, which can be highly problematic~\cite{riehl2024towards} as this can lead to extreme waiting times for few vehicles~\cite{raeis2021deep} and excessive queue forming at feeder roads when prioritizing arterial roads~\cite{gregoire2014capacity}, and generally unequal waiting times~\cite{li2020fairness}.
Moreover, considering fairness is crucial to ensure compliance and acceptance for new control policies by the public~\cite{riehl2024quantitative}.
Therefore, a growing branch of research is dedicated to exploring fairness in the context of traffic control.
Fairness is challenging to study, as there is no consensus on one specific definition of fairness, but generally it is a question of distribution and dianemetic resource allocation problems of spatio-temporal resources in road networks~\cite{riehl2024quantitative,riehl2024towards}.
\textbf{
The few works exploring fair signal control are limited (i) in theoretic models and algorithms with insufficient real-world applicability, (ii) they neglect the multidimensionality fairness and do not systematically analyse fairness, and (iii) benchmarking across established controllers often remains elusive.}
% •	Prior research mainly focused optimized on efficiency measures, which can be highly problematic [riehl2025], leading to extreme waiting times for few [2021_RaeisGarcia] and excessive queue forming at feeder roads to prioritize arterial roads [2014_Gregoire], and highly unequal waiting times. A growing branch of research is therefore dedicated with exploring fairness in the context of traffic control [why, e.g. public acceptance, reasons from riehl2025 et al].
% •	Fairness is hard to define, talk about distributive fairness, dianemetic problem, reference to riehl2025
% •	Related works on fairness-enhancing traffic light control are limited to
% o	Neglecting multidimensionality of fairness & unsystematic analysis of fairness (advocating specific ideologies)
% o	Theoretic models and algorithms with limited real-world applicability (esp. RL methods)
% o	Limited benchmarking, non-systematic across established controllers 

This work sets out to develop a fairness-enhancing design proposition to the most established traffic light control systems SCOOTS and SCATS for a real-world implementation and impact.
The implications of such a controller are systematically explored for a multitude of different fairness and efficiency measures, and a holistic benchmark across various control algorithms is conducted.
The proposed FairSCOSCA method for coordinated, signalized intersection management, is based on two design features at green-phase optimization and early-phase-termination. 
Its feasibility and performance in terms of efficiency and fairness is demonstrated at a real-world, microsimulation case study of the arterial network Schorndorfer Strasse in the city of Esslingen (Germany) covering five intersections. 
The results indicate that fairness can be improved without sacrificing efficiency de trop, and that simple changes to SCOOTS and SCATS can have a significant real-world impact.
A reproducible, open-source implementation can be found on GitHub: \url{https://github.com/DerKevinRiehl/fair_scosca}.
% •	Therefore, this work is dedicated (i) to develop feasible design changes of SCOOTs and SCATs to achieve greater levels of fairness with a great potential for real-world impact, and (ii) to systematically explore the implications of such a control from a multitude of different fairness perspectives, (iii) holistic benchmark across various established control algorithms. We propose FairSCOSCA – a fairness-aware SCOOTs/SCATs traffic light controller based on Bayesian optimization. The feasibility demonstrated and gains in terms of efficiency and equity of the proposed method is analyzed based on the case study of Esslingen, arterial network with seven intersections. The results indicate that fairness can be improved without sacrifice of efficiency de trop, and that simple changes to SCOOTs SCATS could have a significant real-world impact.

The remainder of this work is as follows.
Section~\ref{lab:relworks} reviews related works on fair signal control and identifies gaps in the literature.
Section~\ref{lab:methods} develops the proposed FairSCOSCA method.
Section~\ref{lab:results} presents the case study, and analyses the efficiency and fairness at a benchmark of the proposed and other established controllers.
Section~\ref{lab:discussion} discusses the findings, and Section~\ref{lab:conclusion} summarizes and concludes this work.
% •	The remainder of this work is as follows.

%%%%%%%%%%%%%%%%%%%%%%%%%%%%%%%%%%%%%%%%%%%%%%%%%%%%%%%%%%%%%%%%%%%%%%%%%%%%%%%%
\section{RELATED WORKS} \label{lab:relworks}

% •	Fairness in Traffic
% o	Explain why fairness, explain how like distribute fairness relevant resources, explain different ideologies / perspectives on how to define fairness, and that it is a multidimensional topic with no clear consensus
Fairness plays a significant but largely-overlooked role in the context of traffic control for a multitude of reasons~\cite{riehl2024towards}.
First, it determines the well-being of road users and serves as a hygiene factor for the experience of a drive~\cite{gurney2021equity}.
Second, it enhances road safety through compliance. Control measures to avoid collisions rely on the compliance of road users with signals, but excessively long waiting times are often perceived as unfair, leading to violation of red signals, and accidents~\cite{wissema2002driving}.
Third, public support for investments and implementation of new traffic control measures heavily rely on perceived benefits. Solutions that are perceived as more fair are more likely to be approved by political processes and adapter by the public~\cite{van2022influence}.

There are different ideologies (notions) of how fairness has to be defined, but as no clear consensus has been found in literature, this work uses four notions that seem to suit the current investigation and that are widely spread~\cite{riehl2024towards,riehl2024quantitative}: (i) Egalitarianism, (ii) Rawlsianism, (iii) Utilitarianism, (iv) Harsanyianism.
Egalitarianism~\cite{goppel2016handbuch} considers an equal distribution equitable.
Rawlsianism~\cite{rawls1971atheory} bases on the difference principle and aims to achieve distributions, where the worst-case is improved as much as possible, without necessarily reject disparities.
Utilitarianism~\cite{mill2016utilitarianism} bases on the greater-good principle and argues, that the benefit of the many weighs more than the suffering of the few; according to this principle maximizing efficiency is considered fair.
Harsanyianism~\cite{harsanyi1975can} comprises elements from Rawlsianism and Utilitarianism, and promotes a distribution that achieves best average outcomes.

Works on fair traffic light control in particular explored the distribution of delay and green time durations, and queue lengths~\cite{riehl2024towards}.
% •	Fair Traffic Light Control
% o	Summarize all prior works
% o	Emphasize which fairness ideologies they use and what they are limited in
\textit{Zhang et al. (2010)} suggest dynamically updating min and max green times with a stochastic model, achieving a multi-objective optimization of efficiency and fairness goals~\cite{zhang2010optimizing}. 
Fairness is measured as standard deviation of delay times.
\textit{Slavin et al. (2013)} explore SCOOTS and SCATS in the context of transit signal priority ~\cite{slavin2013statistical}. 
Fairness is implicitly assessed via the average delay of public transport.
\textit{Gregoire et al. (2014)} propose taking queue capacities into account by normalizing pressures to cope with failing back-pressure control in the context of finite road network capacities ~\cite{gregoire2014capacity}. 
Fairness is described by equal queue pressures (lengths).
\textit{Li et al. (2020)} develop a reinforcement learning method for traffic light control dedicated to fairness rather than efficiency using a Fairness Scheduling Proximal Policy Optimization~\cite{li2020fairness}. 
Fairness is captured as the maximum delay any of the vehicles experiences.
\textit{Yan et al. (2020)} extend the learning based approach with a novel reward function to consider an equity factor and adaptive discounting~\cite{yan2020efficiency}. 
Fairness is assessed as standard deviation of travel times.
\textit{Raeis et al. (2021)} elaborate on a deep reinforcement learning method based on two notions of fairness (vehicle and stream)~\cite{raeis2021deep}. 
Fairness is quantified by Jains fairness index, the 95\%-percentile of delays, and volume-proportional measures.
\textit{Huang et al. (2023)} explore a fairness-aware, model-based, multi-agent reinforcement learning method for coordinated traffic signal control~\cite{huang2023fairness}. 
Fairness is measured as coefficient of variation of travel times.
\textit{Shirasaka et al. (2023)} propose a distributed, agent-based, deep reinforcement learning signal controller with fairness~\cite{shirasaka2023distributed}.
Fairness is defined as variation of waiting times.
\textit{Ye et al. (2023)} explore a fairness-aware hierarchical traffic control using a user-satisfaction-index-inspired fairness measure~\cite{ye2022fairlight}.
Fairness is measured as standard deviation of relative delays (delay divided by travel time).

Most of these works share, that they follow an Egalitarian notion of fairness, or that they solely consider one specific ideology of fairness, neglecting the multidimensionality of the equity question.
While the research focus has shifted from single intersection control to coordinated signal control lately, non-learning-based methods, such as adaptions to pressure and SCOOTS/SCATS based methods, have been overlooked by these recent studies. 
Especially a discussion of real-world applicability remains not thematized.
Another limitation that previous works share, is that there is no systematic benchmarking with a variety of established methods, especially non-learning-based methods are often overlooked.

\begin{figure*}[!ht]
    \centering
    \includegraphics[width=\linewidth]{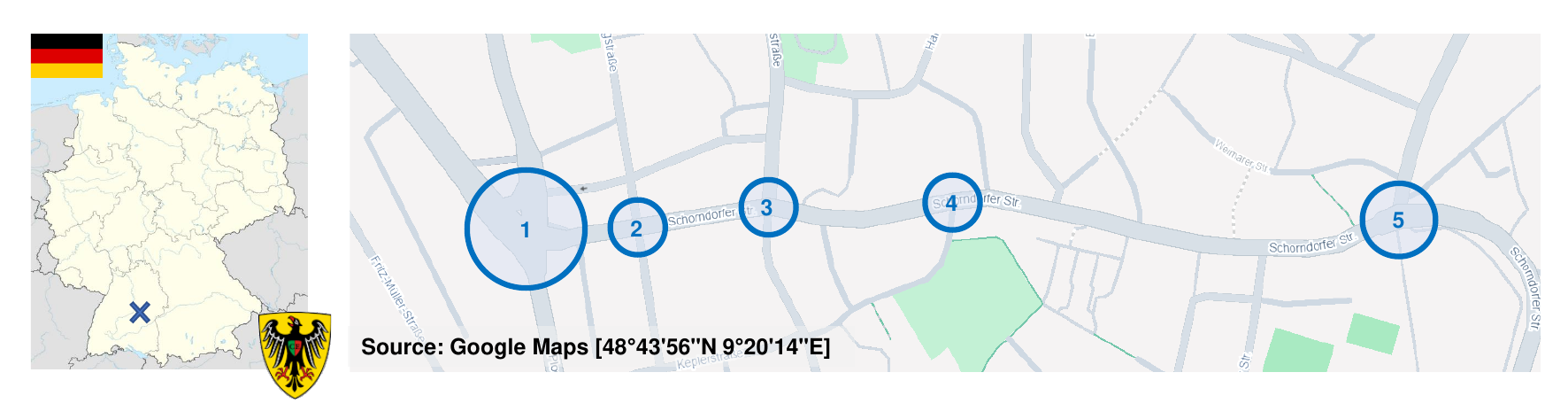}
    \caption{\textbf{Case Study "Schorndorfer Strasse":} The arterial network "Schorndorfer Strasse" with five signalized intersections from the city of Esslingen am Neckar in Germany serves as case study to demonstrate the potential of the proposed FairSCOSCA traffic light controller. The traffic microsimulation model is calibrated based on real-world demand, covers 22 traffic lights, 29 loop detector sensors, 26 bus stops, and 11 bus lines besides car, motorcycle, and truck traffic.}
    \label{fig:casestudy}
\end{figure*}

This work provides a three-fold contribution to the existing literature.
First, the growing branch of coordinated signal control is complemented with a feasible, proposed adaption design feature of SCOTS and SCATS rather than another learning framework.
Second, the present study provides an extensive benchmark of the proposed method with established signal controllers, including pretimed Fixed-Cycle, Max-Pressure, and SCOOTS/SCATS, both in terms of efficiency and fairness.
Third, this study captures the complexity of the equity discussion by assessing fairness from a multi-perspective view.

%%%%%%%%%%%%%%%%%%%%%%%%%%%%%%%%%%%%%%%%%%%%%%%%%%%%%%%%%%%%%%%%%%%%%%%%%%%%%%%%
\section{METHODS}  \label{lab:methods}

\subsection{Notation \& Problem Statement}

The following notation is used to develop a coordinated traffic light control system for arterial networks, that usually consist of a busy arterial road connected to smaller feeder roads with varying demand, and leverages loop-detectors for sensing the system state and three-headed traffic lights (red, yellow, green) for system actuation (provision of way-of right).
The road network is represented as a directed graph $\mathcal{G}$ with nodes $n \in N$, that are intersections, and links $z\in Z$, that connect these intersections: $\mathcal{G} = \{ N, Z\}$.
A set of non-conflicting links $z_j$, that can simultaneously receive right-of-way without collision, forms a movement phase $j$.
The signal control plan of intersection $n$ assigns green durations $g_{n,j}(k)$ for each cycle $k$, based on a fixed set of possible movement phases $j \in F_n$. 
Offsets $O_n(k)$ can be applied at each intersection $n$, representing the delay of the first phase in cycle $k$, in order to create green waves.
The transition from one to the next phase requires a safety-critical period of yellow signals.
The state of each link $z$ is determined by its degree of saturation and the current number of vehicles on this link. 
The degree of saturation $DS_z(k)$ is defined as in SCATS~\cite{SNUG2023}:
\begin{equation}
    DS_z(k) = \frac{g_{n,j}(k)-W_z(k)}{g_{n,j}(k)}
\end{equation}
where $g_j(k)$ denotes  the green time allocated to the phase associated with link $z$ and $W_z$ represents the unused portion of green time on that link. It is computed as:
\begin{equation}
    W_{z}(k) = T_{NO,z}(k)-s_z(k)*T_{OST}
\end{equation}
In this formulation, $T_{NO,z}(k)$ is the total duration (in seconds) during which the stop-line detector on lane $z$ remained unoccupied while the signal was green. 
The term $s_z(k)$ refers to the number of vehicles that passed through the intersection from lane $z$ during the green phase, and $T_{OST}$ is the optimal space time, defined as the inverse of the ideal saturation flow. 
The term $W_z(k)$ thus captures the waste time on lane $z$: the amount of green time that was not utilized but could have been used by additional vehicles under ideal conditions.
The controller's task can be summarized as to define a signal control plan with $g_{n,j}(k)$ given the current state of the network.

\subsection{SCOOTS \& SCATS (SCOSCA)}

SCOOTS and SCATS count amongst the most widely used traffic light control systems.
Despite having slight implementation differences (e.g. sensor data is obtained at different locations on the lanes), they resemble strongly, which is why they are described as SCOSCA in the following.
While the exact implementation details remain part of a business secret, various studies reveal their inner functioning, which is based mainly on three optimization problems~\cite{Stevanovic2009}.
The description of the baseline implementation used in this work follows, which is trying to cover common principles following documentation and academic literature and ~\cite{Lee2002,Chiu1993,McCann2014,slavin2013statistical,Stevanovic2009,SCATS_Home,TRL_SCOOT}.

\textbf{(1) Green Phase Optimization: }
This optimizer assesses the maximum degree of saturation (DS) for each phase $j$ among all the lanes $z$ associated to phase $j$ at an intersection $n$ and calculates the differences between these maxima. 
The phase with the highest DS of the intersection $n$ receives the greatest share of green time, determined by two factors: (i) the difference between the highest and lowest maximum DS values $DS_{diff}(k) = \|max_{z \in j_1}(DS_z) - max_{z\in j_2}(DS_z) \|$, and (ii) an adjustment factor $\lambda_1$, which is determined by Bayesian optimization based on a predefined cost function. 
Bayesian optimization in this context is well-suited to determine control parameters as it models non-convex cost landscapes using Gaussian Processes.
A maximum green time $g_{max}$ cap ensures that no phase is excessively favoured. 
Consecutively, for two phases the following formula would be used to distribute green time (assuming $j$ contains the lane with the highest DS in the intersection):
\begin{equation}
        g_{n,j}(k+1) = \min(g_{max},
        g_{n, j}(k) + DS_{diff}(k) \cdot \lambda_1)
\end{equation}
\begin{equation}
        g_{n,j+1}(k+1) = C(k) - g_{n,j+1}(k)
\end{equation}
where $C(k)$ represents the cycle length.
To prevent unnecessary changes during low traffic conditions, the optimizer includes a
threshold mechanism which is checked via a threshold parameter $\tau_1$.
This mechanism checks whether the number of vehicles on the lane associated with the highest DS exceeds a threshold. If not, the signal timings from the previous optimization step are retained. 
The green phase optimizer is called after each cycle at each intersection. 

\textbf{(2) Cycle Length Optimization: }
The optimizer dynamically adjusts the signal cycle length based on the highest DS among all lanes observed in the network. 
The goal is to maintain this value close to 0.9, as in SCOOTS~\cite{Lee2002,Stevanovic2009}. 
When the maximum DS exceeds 0.925, the cycle length is increased; if it falls below 0.875, the cycle length is reduced~\cite{Lee2002,Stevanovic2009}. 
This hysteresis behaviour follows the Schmitt trigger principle~\cite{Schmitt}, helping to prevent frequent oscillations around the target value. 
Again an adjustment factor $\lambda_2$ is applied when updating the cycle length. 
Additionally the cycle length is updated proportionally to the highest DS $DS_{max}(k)$, similar to SCATS~\cite{Stevanovic2009}. 
To avoid extreme timings, the cycle length $C(k)$ is constrained within predefined minimum ($C_{min}$) and maximum ($C_{max}$) limits. 
When the optimizer is triggered (every fifth cycle), the cycle length $C(k)$ for all intersections is updated according to the following rules: 

\begin{itemize}
    \item If the maximum degree of saturation $DS_{max}$ exceeds 0.925, the cycle length C is increased:
\end{itemize}
\begin{equation}
    C(k+1) = \min(C_{max}, C(k) + (DS_{max}(k) - 0.925) \cdot \lambda_2)
\end{equation}

\begin{itemize}
    \item If $DS_{max}$ falls below 0.875, the cycle length is reduced:
\end{itemize}
\begin{equation}
    C(k+1) = \max(C_{min},C(k) + (0.875 - DS_{\max}(k)\cdot \lambda_2)
\end{equation}

\textbf{(3) Offset Optimization: }
The optimizer adjusts signal offsets $O_n(k)$ between intersections $n$ to form green waves that prioritize the most congested district in the network. 
%The whole network is divided into three districts as shown in ~\ref{fig:Districts}, and 
Congestion is measured by normalizing the total number of vehicles by the number of lanes in each district to ensure fair comparison.
The district with the highest congestion ratio is selected as the critical district, and offsets are applied accordingly:

\begin{itemize}
\item If the district located at the centre of the network (hereafter referred to as the middle district) experiences the highest congestion, apply the following offsets $O_n$ for $N$ intersections:
\end{itemize}
\begin{equation}
    O_{N/2}(k+1) = 0
\end{equation}
\begin{equation}
    O_{n}(k+1) = O_{n+1}(k+1) + \lambda_3 \cdot TT_{n,n+1} \quad \forall n < N/2
\end{equation}
\begin{equation}
    O_{n}(k+1) = O_{n-1}(k+1) + \lambda_3 \cdot TT_{n-1,n} \quad \forall n > N/2
\end{equation}

\begin{itemize}
\item If the critical front (upstream) or back (downstream) districts, assign offsets:
    \begin{itemize}
        \item For reference junction (n = 1 \text{ or } n = N):
    \end{itemize}
\end{itemize}
\begin{equation}
    O_{n}(k+1) = 0
\end{equation}
\begin{itemize}
    \item[] %
    \begin{itemize}
        \item For front intersection:
    \end{itemize}    
\end{itemize}
\begin{equation}
    O_{n}(k+1) = O_{n-1}(k+1) + \lambda_3 \cdot TT_{n-1,n} 
    \quad \forall n > 1
\end{equation}
\begin{itemize}
    \item[] %
    \begin{itemize}
        \item For back intersection:
    \end{itemize}    
\end{itemize}
\begin{equation}
    O_{n}(k+1) = O_{n+1}(k+1) + \lambda_3 \cdot TT_{n,n+1} \quad \forall n < N
\end{equation}
    
where $\lambda_3$ is the adjustment factor and $TT_{n,n+1}$ is the travel time between junction $n$ and junction $n+1$.
Offsets are only applied if the congestion difference between the top two districts exceeds threshold parameter $\tau_2$ (which is again Bayesian Optimized). 
The maximum offset $O_n(k)$ is capped at the cycle length $C(k)$, and the optimizer runs every fifth cycle. 
This method is conceptually similar to SCOOTS use of queue lengths to adjust offsets~\cite{Lee2002,Stevanovic2009}.

\subsection{FairSCOSCA}

We propose two design feature adoptions to SCOSCA: (i) inclusion of cumulative waiting times on opposing phases in the Green Phase Optimization, and (ii) early phase termination as an additional, fairness-enhancing mechanisms to complement the three optimizers.

\textbf{Design Feature FairSCOSCA\_1: }
This feature introduces fairness into the the Green Phase optimization, by adjusting the green time allocation $g_{n,j}(k)$ not only based on the degree of saturation $DS_z(k)$, but also by incorporating the cumulative waiting times of vehicles on opposing phases. 
This dual consideration is intended to improve both the Gini coefficient~\cite{dorfman1979formula} and the maximum delay, thereby addressing fairness from both Egalitarian~\cite{goppel2016handbuch} and Rawlsian~\cite{rawls1971atheory} perspectives.

Unlike SCOSCA that allocates green times purely based on traffic load, this design feature penalizes phases that would otherwise dominate excessively, by factoring in how long vehicles on opposing lanes have been waiting. 
A focus on efficiency only (SCOSCA) systematically penalizes feeder legs with smaller number of vehicles; therefore, the proposed design features aims to find a fair, balanced compromise between feeder and arteria roads, which is acceptable for all users.
For each intersection $n \in N$, the optimizer adjusts $g_{n,j}(k+1)$ based on both the saturation difference $DS_{\text{diff}}(k)$ and a delay-based penalty term $P(k)$ (assuming movement phase $j \in F_n$ has the highest degree of saturation). 
The adjustment balances both components through a tunable parameter 
$\alpha \in [0,1]$.
To capture the severity of waiting times, the penalty term is defined as:
\begin{equation}
    P(k) = e^{N_s(k)} - 1    
\end{equation}
where $N_s(k) = \max(N_z(k)) \;\forall z \in Z$ \text{ such that } $\text{phase}(z) \neq j$ and $N_z$ is the cumulative waiting time of all the vehicles on link $z$. 
The exponential formulation ensures that very high delays are penalized more strongly, while smaller delays have a minimal effect.

The updated green time for the dominant phase j is then computed as:
\begin{equation}
 g_{n,j}(k+1) = \min(g_{max}, max(g_{n,j}(k), g_{n,j}(k) + g_{rw}))
\end{equation}
where $g_{rw}$ can be described by:
\begin{equation}
g_{rw} = \left( \alpha \cdot DS_{diff}(k) - (1 - \alpha) \cdot P(k) \right) \cdot \lambda_1 ) 
\end{equation}

This formulation ensures that green times are increased proportionally to demand $(DS_{\text{diff}}(k)$ while being scaled down if the opposing lanes exhibit high accumulated waiting time $(P(k))$.

\textbf{Design Feature FairSCOSCA\_2: }
%\subsection{SCOSCAFAIRV2 - Add A Secondary Fairness Controller}
This feature retains the three SCOSCA optimizers for green phases, cycle lengths, and offsets, but augments it with a secondary fairness-aware mechanism - early phase termination. 
This additional controller introduces short-term fairness interventions by adjusting the phase durations in response to unfavourable conditions for individual vehicles.
The mechanism is triggered when a vehicle arrives at a red signal during the early portion of the currently active phase. 
Let $j_a \in F_n$ be the currently active movement phase at intersection $n \in N$, and let $j_r \in F_n$ be the phase corresponding to the red approach where the vehicle has arrived. 
If the active phase $j_a$ will be still active for more than a tunable threshold called \texttt{TTG} (time to green) (indicating the time the vehicle still has to wait), the controller allows a pre-emption to take place.

In this case, the green time allocation for the active phase $g_{n,j_a}(k)$ is shortened by a fixed duration \texttt{TEG} (time earlier green), and the green time for the waiting phase $g_{n,j_r}(k)$ is increased by the same amount. 
This operation preserves the total cycle length:
\begin{equation}
    g_{n,j_a}(k) \leftarrow g_{n,j_a}(k) - \texttt{TEG}, \quad    
\end{equation}
\begin{equation}
    g_{n,j_r}(k) \leftarrow g_{n,j_r}(k) + \texttt{TEG}  
\end{equation}

To address potential long-term imbalances, this feature applies a compensation mechanism in the subsequent cycle. 
The phase $j_a$, which was previously shortened, receives additional green time equal to the amount it lost:
\begin{equation}
    g_{n,j_a}(k+1) \leftarrow g_{n,j_a}(k+1) + \texttt{TEG}
\end{equation}
To prevent instability caused by too many adaptations, at most one such pre-emption is permitted per junction per cycle. 
Furthermore, if a pre-emption occurs at a junction in cycle $k$, no further pre-emptions are allowed at that junction in cycle $k+1$, thereby enforcing the compensation.

This method aims to reduce extreme delays for vehicles that would otherwise wait quite long, having just missed a green light. 
By reducing extreme delays and waiting times, this feature thus aligns conceptually with Rawlsian fairness and Egalitarian fairness, and introduces two parameters \texttt{TTG} and \texttt{TEG}.

\begin{figure*}[!ht]
    \centering
    \includegraphics[width=0.9\linewidth]{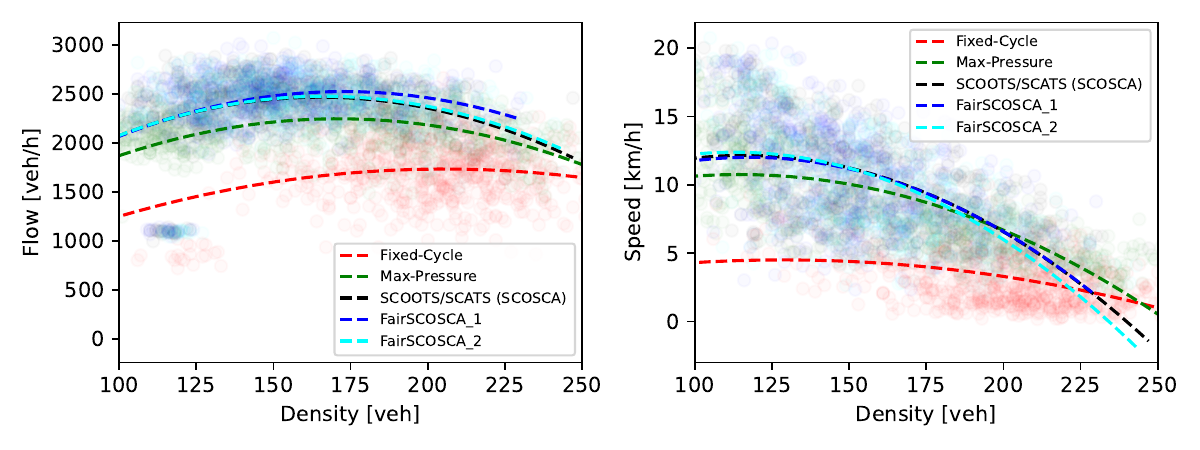}
    \caption{\textbf{Efficiency Analysis:} Visualization of Macroscopic Fundamental Diagram (MFD). The two figures show flow and average vehicle speed for varying vehicle density across the simulation. The Fixed-Cycle controller obtains the lowest network capacity, Max-Pressure gains improvements. The proposed design features FairSCOSCA\_1 and FairSCOSCA\_2 obtain even further efficiency gains, similar to those of SCOOTS/SCATS (SCOSCA) (best). The MFD was calculated for every 5 minutes across all simulation runs (with different seeds), and the density is measured as the number of vehicles in the case study network.}
    \label{fig:efficiency_analysis}
\end{figure*}

\begin{figure*}[!ht]
    \centering
    \includegraphics[width=0.9\linewidth]{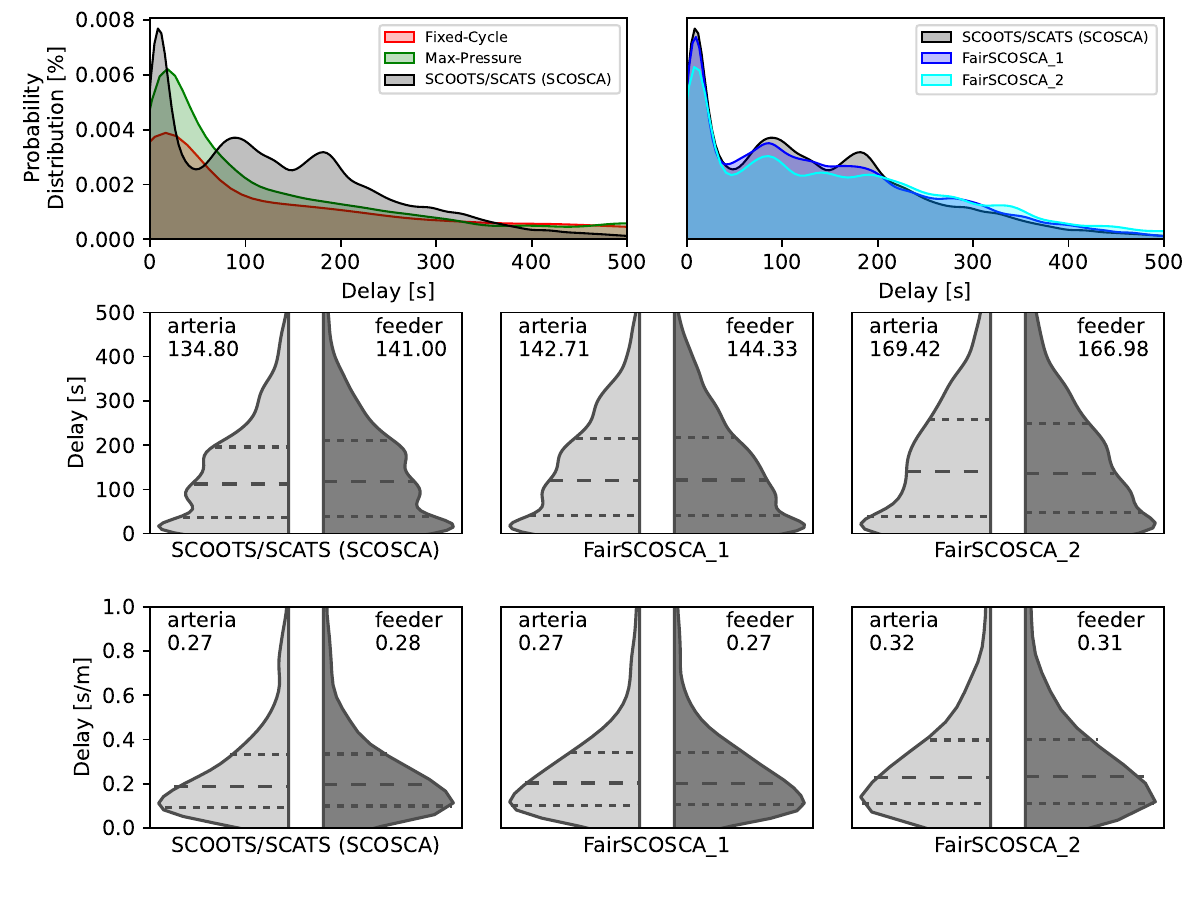}
    \caption{\textbf{Equity Analysis:} The diagrams of the first row show the distribution of delays for different controllers.
    Contrary to Fixed-Cycle and Max-Pressure, SCOOTS/SCATS (SCOSCA) achieves lower average delays and a higher concentration of those delays (more equal distribution). The proposed design features achieve similar results with slight deviations.
    The diagrams of the second two rows show the delay and normalized distribution for vehicles starting their trip from arterial or feeder roads for the different versions of SCOSCA. Normalized delays refer to delay per travelled distance.
    The number below "arteria" and "feeder" describe the median of the delay.
    While vehicles from the arterial roads have lower delays for SCOSCA (discrimination of feeders), the proposed design features enable to improve the situation of feeders without impairing those vehicles from the arteria.
    }
    \label{fig:equity_analysis}
\end{figure*}

\begin{table*}[!ht]
    \caption{\textbf{Benchmark of Traffic Controllers:} This table compares the investigated traffic light controllers in terms of efficiency (A), equity/fairness (B), and horizontal equity between arteria and feeder roads (C). Each value is reported as mean across 20 traffic microsimulations, followed by standard deviation in rectangular brackets, and a statistical significance symbol. The statistical significance symbol indicates whether the controller was performing significantly higher (+) or lower (-) compared to the baseline SCOSCA at the following p-values: 1\% (+++), 2\% (++), and 5\% (+). Green indicates improvements, while red denotes impairments.}
    \label{tab:results}
    \centering
    \begin{tabular}{lllll}
        \textbf{(A) Efficiency Measures} \\
        \textbf{Method} & \textbf{Flow [veh/h]} & \textbf{Speed [m/s]} & \textbf{Density [veh]} & \textbf{Throughput [veh]} \\ 
        \hline
        SCOSCA & 4901.75 [128.02] & 2.52 [0.26] & 131.20 [11.21] & 12134.90 [389.94] \\
        FairSCOSCA\_1 & 5018.70 [85.54] \textcolor{green}{+++} & 2.68 [0.22] \textcolor{green}{+} & 125.23 [10.70] & 12470.35 [241.29] \textcolor{green}{+++} \\
        FairSCOSCA\_2 & 4924.50 [91.52] & 2.49 [0.19] & 132.48 [8.54] & 12191.05 [264.47] \\ 
        Max-Pressure & 4233.20 [37.35] \textcolor{red}{- - -} & 1.99 [0.11] \textcolor{red}{- - -} & 140.00 [6.57] \textcolor{red}{+++} & 10437.90 [122.72] \textcolor{red}{- - -} \\ 
        Fixed-Cycle & 3487.95 [99.56] \textcolor{red}{- - -} & 1.19 [0.09] \textcolor{red}{- - -} & 166.62 [5.16] \textcolor{red}{+++} & 7462.20 [283.91] \textcolor{red}{- - -} \\ 
        \\
        \textbf{(B) Equity Measures} \\
         & (Egalitarian) & (Rawlsian) & (Utilitarian) & (Harsanyian) \\ 
        \textbf{Method} & \textbf{Gini of Delays} & \textbf{Max. Delay [s]} & \textbf{Total Travel Time [h]} & \textbf{Avg. Delay [s]} \\ 
        \hline
        SCOSCA & 0.5085 [0.02] & 877.90 [322.85] & 326.04 [26.05] & 127.28 [21.48] \\
        FairSCOSCA\_1 & 0.4993 [0.01] \textcolor{green}{-} & 852.30 [268.03] \textcolor{green}{-} & 313.94 [25.05] \textcolor{green}{-} & 114.28 [16.52] \textcolor{green}{-} \\
        FairSCOSCA\_2 & 0.4955 [0.01] \textcolor{green}{-} & 808.15 [183.93] \textcolor{green}{- -} & 330.80 [20.72] & 128.58 [16.16] \\ 
        Max-Pressure & 0.5773 [0.01] \textcolor{red}{+++} & 1247.35 [572.05] \textcolor{red}{+} & 329.11 [16.39] \textcolor{red}{+} & 145.12 [10.75]\textcolor{red}{+++} \\ 
        Fixed-Cycle & 0.6414 [0.02] \textcolor{red}{+++} & 2676.60 [710.07] \textcolor{red}{+++} & 410.87 [15.62] \textcolor{red}{+++} & 324.76 [24.95] \textcolor{red}{+++} \\ 
        \\
        \textbf{(C) Horizontal Equity} \\
        \textbf{Method} & \textbf{Md. Delay Arteria [s]} & \textbf{Md. Delay Feeder [s]} & \textbf{Gini Arteria} & \textbf{Gini Feeder} \\ 
        \hline
        SCOSCA  &115.41 [23.54] & 137.70 [19.93] & 0.4588 [0.02] & 0.5632 [0.03] \\
        FairSCOSCA\_1 & 97.98 [19.22] \textcolor{green}{- -} & 128.86 [16.33]  & 0.4553 [0.02] & 0.5435 [0.01] \textcolor{green}{- - -} \\
        FairSCOSCA\_2 & 115.03 [16.37] & 140.49 [16.30] & 0.4434 [0.01] \textcolor{green}{- -} & 0.5532 [0.02] \\
        Max-Pressure & 132.68 [10.10] & 163.97 [12.76] \textcolor{red}{+++} & 0.5228 [0.01] \textcolor{red}{+++} & 0.6284 [0.01] \textcolor{red}{+++} \\
        Fixed-Cycle & 318.24 [29.90] \textcolor{red}{+++} & 337.58 [22.52] \textcolor{red}{+++} & 0.5809 [0.02] \textcolor{red}{+++} & 0.7332 [0.03] \textcolor{red}{+++} \\
        \\
        \hline
    \end{tabular}
\end{table*}

\begin{table*}[!ht]
    \caption{\textbf{Detailed Horizontal Equity Analysis:} This table assesses the discrimination of vehicles originating from arteria and feeder roads from various fairness metrics. Each value is reported as mean across 20 traffic microsimulations.
    The analysis was conducted for delays (A) and delays per travelled distance (B). For the second analysis, a Utilitarian analysis was excluded as it does not make sense in this context.}
    \label{tab:results2}
    \centering
    %\resizebox{0.9\textwidth}{!}{
    \begin{tabular}{lrrrrrrrr}
        \textbf{(A) Delay [s]}
         & 
        \multicolumn{2}{c}{(Egalitarian)} & 
        \multicolumn{2}{c}{(Rawlsian)} & 
        \multicolumn{2}{c}{(Utilitarian)} & 
        \multicolumn{2}{c}{(Harsanyian)} \\
         & 
        \multicolumn{2}{c}{\textbf{Gini of Delay}} & 
        \multicolumn{2}{c}{\textbf{Max. Delay [s]}} & 
        \multicolumn{2}{c}{\textbf{Total Delay [h]}} & 
        \multicolumn{2}{c}{\textbf{Avg.Delay [s]}} \\
        \textbf{Method} & Arteria & Feeder & Arteria & Feeder & Arteria & Feeder & Arteria & Feeder \\
        \hline
        SCOOTS/SCATS & 0.4588 & 0.5632 & 661.38 & 638.53 & 107.32 & 126.62 & 134.80 & 140.99 \\
        FairSCOSCA\_1 & 0.4553 & 0.5435 & 606.79 & 631.36 & 114.05 & 130.10 & 142.71 & 144.33 \\
        FairSCOSCA\_2 & 0.4434 & 0.5532 & 1022.49 & 1028.43 & 132.39 & 147.22 & 169.42 & 166.98 \\
        Max-Pressure & 0.5228 & 0.6284 & 1492.98 & 1298.39 & 126.04 & 146.09 & 183.19 & 187.63 \\
        Fixed-Cycle & 0.5809 & 0.7332 & 2024.82 & 2042.98 & 176.62 & 209.28 & 312.61 & 328.00 \\
        \\
        \\
        \textbf{(B) Delay [s/km]}
         & 
        \multicolumn{2}{c}{(Egalitarian)} & 
        \multicolumn{2}{c}{(Rawlsian)} & 
        \multicolumn{2}{c}{(Utilitarian)} & 
        \multicolumn{2}{c}{(Harsanyian)} \\
         & 
        \multicolumn{2}{c}{\textbf{Gini of Delay}} & 
        \multicolumn{2}{c}{\textbf{Max. Delay}} & 
        \multicolumn{2}{c}{\textbf{Total Delay}} & 
        \multicolumn{2}{c}{\textbf{Avg.Delay}} \\
        \textbf{Method} & Arteria & Feeder & Arteria & Feeder & Arteria & Feeder & Arteria & Feeder \\
        \hline
        SCOOTS/SCATS   & 0.4616 & 0.4584 & 2739 & 5052 & - & - & 266.87 & 275.05 \\
        FairSCOSCA\_1  & 0.4488 & 0.4417 & 3423 & 3197 & - & - & 274.59 & 272.62 \\
        FairSCOSCA\_2  & 0.4714 & 0.4617 & 6429 & 7973 & - & - & 317.98 & 313.45 \\
        Max-Pressure   & 0.5051 & 0.5104 & 3432 & 3231 & - & - & 324.41 & 337.66 \\
        Fixed-Cycle    & 0.5443 & 0.5404 & 7175 & 7184 & - & - & 534.25 & 534.73 \\
        \\
        \hline
    \end{tabular}
    %}
\end{table*}

%%%%%%%%%%%%%%%%%%%%%%%%%%%%%%%%%%%%%%%%%%%%%%%%%%%%%%%%%%%%%%%%%%%%%%%%%%%%%%%%
\section{RESULTS} \label{lab:results}

\subsection{Benchmark Microsimulation Case Study}
The proposed method is evaluated using a traffic microsimulation case study of a demand-calibrated, real-world arterial network of the Schorndorfer Strasse in Esslingen am Neckar (Germany) as shown in Fig.~\ref{fig:casestudy}, originating from a cooperation with the municipal administration Tiefbauamt Esslingen.
The network comprises five signalized intersections, 26 bus stops, 29 sensors, and 22 traffic lights.
The Python implementation of the microsimulation employs the microsimulation environment SUMO~\cite{krajzewicz2002sumo}, and runs on traffic data for a usual afternoon peak working day (March 4th, 2024).
The mixed, multimodal vehicle fleet composition was adjusted according to statistics from the federal office of transport of Germany (Kraftfahrt-Bundesamt), and public transport time tables for 11 bus lines were considered~\cite{fahrzeugbestand}.
Bayesian optimization and 20 random seeds of the microsimulation were used to determine the various control parameters, where the optimization goal was to minimize the overall average delay (maximize efficiency) of the network, due to a highly stochastic, non-linear optimization topology.

\subsection{Efficiency Analysis}
To evaluate the efficiency of SCOSCA and the two fairness-oriented designs, four performance metrics are considered: throughput, flow, density, and average speed. These results are directly compared against two baseline controllers, fixed-cycle and max-pressure, to provide a reference for assessing relative performance.

The macroscopic fundamental diagram in Fig.~\ref{fig:efficiency_analysis} visualizes the network capacity for different controllers. 
The \textit{Fixed-Cycle} controller obtains the lowest network capacity, and Max-Pressure gains improvements. The proposed design features \textit{FairSCOSCA\_1} and \textit{FairSCOSCA\_2} obtain even further efficiency gains, similar to those of \textit{SCOOTS/SCATS} (\textit{SCOSCA}, most efficient).
With regards to a more quantitative assessment, Table~\ref{tab:results}(A) reveals flow, speed, density and throughput improvements.
\textit{FairSCOSCA\_1} is able to achieve significantly higher levels of efficiency when compared with \textit{SCOSCA}, due to its ability to optimize green phases more efficiently leveraging the additional information of cumulative waiting times.
The flow increased by 2.39\%, speed by 6.35\%, and throughput by 2.77\%.
\textit{FairSCOSCA\_2} achieves similar levels of efficiency to \textit{SCOSCA}, with no significant differences.

\subsection{Fairness Analysis}
To evaluate the fairness of the proposed fairness-enhanced control strategies, four metrics are used, each grounded in a distinct fairness ideology. This multidimensional evaluation enables an ideology-independent assessment of fairness performance.
The Gini coefficient corresponds to Egalitarianism and measures how equally delays are distributed among all vehicles. 
The maximum delay reflects Rawlsian fairness, which prioritizes improving outcomes for the worst-off. 
The average delay captures Harsanyian fairness, which considers the mean utility across all users. 
Lastly, the total travel time represents the Utilitarian perspective, which focuses on maximizing system-level utility, potentially at the expense of individual fairness.
%The Gini coefficient and average delay were measured separately for all vehicles, for vehicles originating from the main road, and for those from the side roads, in order to better understand the differences in performance between main and side road traffic.
Besides, a horizontal fairness analysis is conducted, comparing average delay and Gini coefficient of delays for vehicles starting their trip from arteria (main road) or feeder (side road), to assess origin-based  discrimination.

The probability distribution functions of delays are visualized for different controllers in the first row of Fig.~\ref{fig:equity_analysis}, which allow a discussion of Harsanyian and Egalitarian fairness.
\textit{SCOSCA} achieves significantly higher concentration at low delays (resulting in lower Gini coefficient due to a more concentrated distribution) and lower average delays, when compared with \textit{Max-Pressure} (second best) and Fixed-Cycle (worst).
The proposed design features \textit{FairSCOSCA\_1} and \textit{FairSCOSCA\_2} achieve similar results to \textit{SCOSCA}, with slight differences in the delay distribution and little lower concentration at low delays.

Table~\ref{tab:results}(B) comprehensively reveals a systematic comparison of fairness measures from all four notions of fairness.
Contrary to \textit{SCOSCA}, the benchmark controllers \textit{Max-Pressure} and \textit{Fixed-Cycle} perform significantly worse in Egalitarian fairness (higher Gini coefficients), Rawlsian fairness (higher maximum delay), Utilitarian fairness (total travel time), and Harsanyian fairness (average delay).
\textit{FairSCOSCA\_1} achieves significant equity improvements across all fairness measures.
\textit{FairSCOSCA\_2} achieves significant improvements in terms of Egalitarian and Rawlsian fairness only. 

The probability distribution functions of delays for arteria and feeder originating vehicles in the second row of Fig.~\ref{fig:equity_analysis} allow a discussion of horizontal fairness.
While the distribution for arteria-originating users remains similar across the methods, improvements for users originating from feeder roads can be observed for \textit{FairSCOSCA\_1}.
Table~\ref{tab:results}(C) supports the initial assessment. 
\textit{FairSCOSCA\_1} achieves improvements for feeder-originating vehicles in terms of average delay and Gini coefficient, resulting in a smaller discrimination of feeder-originating vehicles.

More specifically, Table~\ref{tab:results2} deep-dives into the horizontal equity analysis, by providing a systematic assessment of all fairness notions differentiating the origin of vehicles.
Across all controllers one can observe a discrimination of feeder-originating vehicles in terms of delays for all fairness notions.
In terms of normed delays (per km) however results are mixed. 
The proposed design feature \textit{FairSCOSCA\_1} achieves a reduction of the discrimination (differences between values) of feeder-originating vehicles.
On the contrary, the proposed design feature \textit{FairSCOSCA\_2} causes a discrimination of arteria-originating vehicles.

%%%%%%%%%%%%%%%%%%%%%%%%%%%%%%%%%%%%%%%%%%%%%%%%%%%%%%%%%%%%%%%%%%%%%%%%%%%%%%%%
\section{DISCUSSION} \label{lab:discussion}

The benchmark shows, that \textit{SCOSCA} itself is more efficient and equitable when compared to \textit{Max-Pressure} and \textit{Fixed-Cycle} control already.
\textit{SCOSCA} primarily relies on the degree of saturation (DS), along with queue length and vehicle density, to inform its control decisions. 
Even though it does not include a dedicated, explicit, fairness-enhancing mechanisms, the observation suggests that \textit{SCOSCA}, by design, inherently promotes a certain level of fairness.
The fairness-promoting property of \textit{SCOSCA} stems from the nature of the DS metric itself. 
Because DS is weighted by the amount of allocated green time, phases that already receive substantial green time tend to exhibit lower DS values. 
This built-in feedback mechanism naturally prevents over-allocating green time to already-favoured phases, thus avoiding extreme unfairness scenarios. 
In contrast, algorithms such as Max-Pressure, which rely purely on instantaneous queue lengths or flow differences, can lead to highly inequitable outcomes where certain phases are almost entirely neglected.

Furthermore, the results highlight, that the proposed design features can achieve significant gains and substantial improvements in equity with only marginal impairments in efficiency.
As such, the proposed method is demonstrated to have a strong potential to fairness-enhance \textit{SCOSCA}.
\textit{FairSCOSCA\_1} performed notably better than \textit{SCOSCA}. 
Not only were fairness metrics improved, but efficiency metrics also benefited. 
This improvement may be attributed to the principle of using more information than just the degree of saturation.
By respecting vehicle waiting times in its green phase optimizer, it can achieve more sophisticated green distributions.
\textit{FairSCOSCA\_2} also showed consistent performance improvements across many metrics. 
The addition of a secondary controller generally enhanced fairness metrics, while resulting trade-offs were only minor. 

The discrimination of feeder-originating vehicles was observed in the horizontal equity analysis and confirms the prior findings of \textit{Gregoire et al. (2014)}~\cite{gregoire2014capacity}.
Especially \textit{Max-Pressure} strongly favours arterial roads, while feeder roads are excessively neglected. 
This results in a large Gini coefficients and maximum delays that arise on the feeder roads. 
\textit{FairSCOSCA\_1} achieves a reduced discrimination, while \textit{FairSCOSCA\_2} even starts discriminating arteria-originating vehicles (in terms of average delays).

Most importantly, the proposed design features are simple, feasible changes to the widely established \textit{SCOOTS/SCATS} traffic light control.
Due to its real-world applicability, the proposed method therefore has a huge real-world potential to improve the most widely adapted traffic control system.
Despite the widespread deployment of \textit{SCOSCA}, fairness has remained largely absent from their design logic. This absence can be attributed to (i) the historical emphasis on efficiency in traffic control research and practice, (ii) the lack of established frameworks to operationalize fairness in traffic systems, and (iii) the proprietary and opaque nature of many existing systems, which has discouraged practical design interventions. 
\textit{FairSCOSCA} addresses this gap by proposing transparent, easily implementable fairness mechanisms, grounded in multiple normative definitions and validated through open-source simulation.

\section{CONCLUSION} \label{lab:conclusion}
This paper set out to propose simple design features to enhance the fairness of the most widely adopted traffic light control systems \textit{SCOOTS} and \textit{SCATS} in the context of arterial networks.
The proposed method \textit{FairSCOSCA} was evaluated based on a realistic, demand-calibrated microsimulation in Esslingen am Neckar (Germany).
The benchmark results highlight significant fairness gains that could be achieved without severe sacrifices in efficiency, from a multitude of fairness definitions.
%•	We proposed a method to fairness-enhance the most widely adapted traffic light control system SCOOTs SCATs and was evaluated based on a realistic, calibrated, microsimulation of arterial network case study of Esslingen in Germany.
%•	The results showed that significant fairness gains could be achieved without great sacrifice of efficiency.

\textbf{Ideas For Future Works:}
\begin{itemize}
    \item This work leveraged a traffic microsimulation for the evaluation of the proposed method. Future research should compare the obtained results in real-world deployments of \textit{SCOOTS} and \textit{SCATS} (SCOSCA) using floating car data. 
    \item Moreover driver perception and acceptance surveys could complement such deployments.
    \item Besides, this work optimized for efficiency only (average delay); an multi-objective optimization for efficiency and fairness measures might help to achieve equity gains with a pure \textit{SCOSCA} controller and would be another interesting direction of research.
    \item What's more, it would be interesting to test the method on more diverse urban networks (e.g., grid layouts, suburban sprawls, multimodal intersections) to assess transferability and robustness beyond the scope of arterial networks, such as the Schorndorfer Strasse case study.
    \item In addition to that, an assessment of robustness under incident conditions.
    Future works should analyze how \textit{FairSCOSCA} and other traffic light controllers perform under incidents, sensor failures, or unplanned events, and whether fairness properties degrade gracefully or disproportionately under stress.
    \item Also, future works could explore how \textit{SCOSCA} could be modified to incorporate prioritization of specific modes, such as public transportation or cycling / active modes, to achieve further fairness-relevant, societal goals.
\end{itemize}

\textbf{Limitations of this Work:}
\begin{itemize}
    \item The study models fairness using delay-based metrics aligned with well-established normative frameworks (e.g., Gini, Rawlsian). However, fairness perceptions may also involve socio-demographics, mode of transport, or purpose of travel—factors not captured here.
    \item The model assumes static or pre-defined demand profiles for the simulation. Dynamic, real-time variations in traffic inflow (e.g., due to weather or events) could affect controller performance and fairness dynamics.
    \item Findings are based on one arterial network in a specific urban context. While this is a meaningful real-world case, broader generalizability to other cities or traffic conditions remains to be tested.
\end{itemize}

%%%%%%%%%%%%%%%%%%%%%%%%%%%%%%%%%%%%%%%%%%%%%%%%%%%%%%%%%%%%%%%%%%%%%%%%%%%%%%%%
\newpage

\bibliographystyle{IEEEtranDOI}
\bibliography{references}

\end{document}